# Class-Based Interference Management in Wireless Networks

Mohammad Arif Hossain[1], Mostafa Zaman Chowdhury[2], Shakil Ahmed[2], and Yeong Min Jang[1]
[1]Dept. of Electronics Engineering, Kookmin University, Seoul 136-702, Korea
[2]Dept. of Electrical and Electronic Engineering, Khulna University of Engineering & Technology, Khulna-9203, Bangladesh
E-mail: dihan.kuet@gmail.com, mzceee@yahoo.com, shakileee076@gmail.com, yjang@kookmin.ac.kr

*Abstract*—Technological advancement has brought revolutionary change in the converged wireless networks. Due to the existence of different types of traffic, provisioning of Quality of Service (QoS) becomes a challenge in the wireless networks. In case of a congested network, resource allocation has emerged as an effective way to provide the excessive users with desirable QoS. Since QoS for non-real-time traffic are not as strict as for real-time traffic, the unoccupied channels of the adjacent cells can be assigned to the non-real-time traffic to retain QoS for real-time traffic. This results in the intensified bandwidth utilization as well as less interference for the real-time traffic. In this paper, we propose an effective radio resource management scheme that relies on the dynamically assigned bandwidth allocation process. In case of interference management, we classify the traffic into real-time traffic and non-real-time traffic and give priority to the real-time traffic. According to our scheme, the real-time traffic among the excessive number of users are reassigned to the original channels which have been occupied by non-real-time traffic and the non-real-time traffic are allocated to the assigned channels of those real-time traffic. The architecture allows improved signal to interference plus noise ratio (SINR) for real-time traffic along with intensification in the bandwidth utilization of the network. Besides, the increased system capacity and lower outage probability of the network bear the significance of the proposed scheme.

## I. Introduction

The world sees the dramatic changes in the field of wireless communication. As radio resource is limited, the management of this radio resource should be done in such way that ensures the standard level of Quality of Service (QoS). Again the limited bandwidth and huge number of users force the technician to assign the unused channels of the nearby cells to the cell where traffic intensity is higher. The interference is the biggest hindrance in this approach. There are real-time and non-real-time traffic in the networks. The real-time traffic should be given higher priority compared to the non-real-time traffic for retaining the highest QoS level. So, steps must be taken that guarantee the standard level of QoS for the majority of the real-time traffic users. If a cell has the number of users which is greater than the total number of channels of the cell, the cell can take the advantage by assigning the unused channels from adjacent cells if there are unoccupied channels in the adjacent cells [1]-[3].

In this paper, an interference management process with priority given to the real-time traffic has been proposed since the QoS for the non-real-time traffic is not very stringent compared to real-time traffic. The proposal is mainly based on dynamic channel assigning scheme. In the scheme, cell having highest traffic intensity is bifurcated in such a way that the same frequency band of adjacent cells can be resided for the inner part user of the cell. This approach diminishes the interference problem when the unused channels of adjacent cells are allocated.

According to the scheme, if new traffic arrives in a cell and all the channels of the cell are fulfilled by the preexisting traffic in the cell, the unused channels of the adjacent cells can be assigned for meeting the demand of the excessive number of users. The cell is then bifurcated and the traffic is assigned to the inner part of the cell to reduce interference. As the borrowed channels can cause interference from the nearby cells, the original channels of the cell are provided to the outer part and the borrowed channels are assigned to the inner part. If the traffic is real-time traffic, the original channels of the cell is provided for it. Consequently, the borrowed channels are provided to non-real-time traffic.

In the previous time, dynamic channel allocation with interference mitigation architecture [4] and interference declination approach for OFDMA networks [5], [6] are described with some interference management techniques. The dynamic channel borrowing scheme and radio resource management scheme [7]-[9] show better and efficient performances for interference management but the models are not concern about the interference management for real-time traffic. However, the proposal of this paper shows the performances considering the priority based QoS for real-time traffic which is not considered in the above models.

The rest of the paper is systematized as followed: Section II shows the proposed interference declination scheme for real-time traffic with proper illustration. The outage probability and system capacity analysis for the proposed scheme are shown in Section III. Section IV demonstrates the simulation results for the proposed scheme and compares the proposed model to the model where real-time traffic and non-real-time traffic is not considered. Finally, conclusions are drawn.

## II. DYNAMIC CHANNEL ASSIGNMENT WITH INTERFERENCE MITIGATION FOR REAL-TIME TRAFFIC

Contemporary wireless network is committed to assist QoS to its consumers. Interference mitigation for real-time traffic is one of the ways to offer standard level of QoS as real-time traffic users are unwilling to face buffering. We propose a dynamic channel assigning scheme with interference mitigation process for ensuring higher level of QoS for real-time traffic. We also show the interference management procedure for non-real-time traffic.

*A. System model*

In this system model, we propose a cluster of seven cells that have frequency band $X$, $A$, and $B$. Suffix used in the frequency band $X$, $A$, and $B$ represents the cell number. We define the cell as reference cell which has highest traffic intensity compared to the other cells in the cluster. In our model, the cell containing $X_1$ (cell 1) is the reference cell. The rest of the cells of the cluster are grouped into two, categorized as $A_2$, $A_4$, $A_6$ and $B_3$, $B_5$, $B_7$. It is assumed that $|A_2|=|A_4|=|A_6|=|B_3|=|B_5|=|B_6|$. The cells which have maximum number of unused channels available in the cells of the two groups are assumed as cell 6 and cell 7. Let, the cell 6 and the cell 7 contains $A_6'$ and $B_7'$ unoccupied channels, respectively, where $A_6'>B_7'$.

Consequently, there arises a certainty to assign unused channels from the adjacent cells to the reference cell if the cell cannot accommodate the excessive number of traffic within its limited number of channels. If $A_6'$ and $B_7'$ channels are assigned to the reference cell (cell 1), then the cell will experience severe interference from $A_2$, $B_3$, $A_4$, and $B_5$. When channels are assigned from cell 6 and cell 7, they have to assign in the inner part of the reference cell to reduce interference. Figure 1 demonstrates that both the real-time and non-real-time traffic may exist in the inner part and outer part of cell 1. The bifurcation process of the reference cell is shown in Fig 1.

Figure 2 shows the procedure to minimize the interference giving priority for real-time traffic. The assigned channels for which interference will be occurred i.e. $A_6'$ and $B_7'$ are provided to the inner part users and outer part has the non-interfering frequency band $X_1$. It is a matter of concern that the real-time traffic should have less interference compared to non-real-time traffic. As the traffic of the inner part in the reference cell then gets interference from the adjacent cells, the real-time traffic of that part should be reassigned with the original channels of the cell. However, the original channels of the outer part belong only to the non-real-time traffic of the reference cell can be assigned to those real-time traffic of the inner part. Consequently, the non-real-time traffic of the original channels is assigned with the released channels of the real-time traffic of the inner part. If there is no non-real-time traffic in the outer part, then real-time traffic is housed by the cell bifurcation processes. Besides, the non-real-time traffic of the inner part is assigned with unoccupied channels of the adjacent cells and the interference is reduced by cell bifurcation.

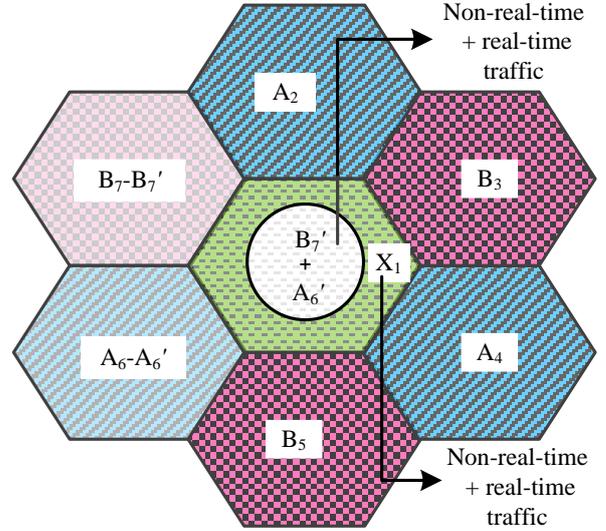

**Fig. 1.** Dynamic channel assigning process.

Figure 2(a) shows that $X_1'$ and $X_1''$ bands (part of $X_1$) are assigned to the real-time traffic of the outer part and the inner part of the reference cell, respectively. $X_1'''$ (part of $X_1$) band is allocated only for the non-real-time traffic of the outer part. Moreover, $A_6''$ (part of $A_6'$) and $B_7''$ (part of $B_7'$) are assigned to the non-real-time traffic of the inner part of the reference cell. Consequently, $A_6'''$ ($A_6'''= A_6'-A_6''$) and $B_7'''$ ($B_7'''= B_7'-B_7''$) bands are provide to the non-real-time traffic of the outer part. As the same frequency bands of the adjacent cell are assigned to the non-real time traffic in the outer part of the reference cell, the interference for the traffic will degrade the QoS severely. Concerning about this problem the adjacent cells are also bifurcated and the interfering channels are provided to the inner part of the respective adjacent cells. Here, $A_2'''$ and $A_4'''$ are the same frequency bands of $A_6'''$ have to kept in the inner part of cell 2 and cell 4, respectively, to reduce the interference in the reference cell. Similarly, $B_3'''$ and $B_5'''$ are the same frequency band of $B_7'''$ have to put in the inner part of cell 3 and cell 5, respectively, to reduce the interference in the reference cell. Thus the interference becomes trifling for real-time traffic and in considerable level for non-real-time.

If the interfering channels of the adjacent cells are unoccupied, then the bifurcation of the reference cell and adjacent cells can be omitted by keeping these channels blocked which is shown in Fig. 2(b). The original frequency band $X_1$ is divided into $X_1'$ and $X_1''$ bands. $X_1'$ and $X_1''$ bands are assigned to the real-time traffic and non-real-time traffic, respectively. $A_6'$ and $B_7'$ bands are provided to the non-real-time traffic. If the interfering channels of the adjacent cell are partially occupied, then the reference cell and other adjacent cells are bifurcated and only the interfering occupied channels are provided to the inner part and the unoccupied channels are kept blocked that is shown in Fig 2(c). Here, $0 \leq A_2'' \leq A_2'''$ and $0 \leq A_4'' \leq A_4'''$. Similarly, $0 \leq B_3'' \leq B_3'''$ and $0 \leq B_5'' \leq B_5'''$. So, it is very easy to understood that if the unoccupied channels $A_2'''$, $A_4'''$, $B_3'''$, and $B_5'''$ become occupied, those channels are delivered to the inner part users of the respective adjacent cells.

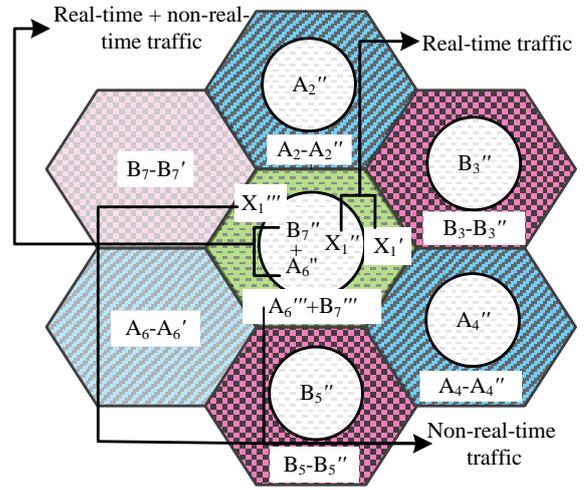

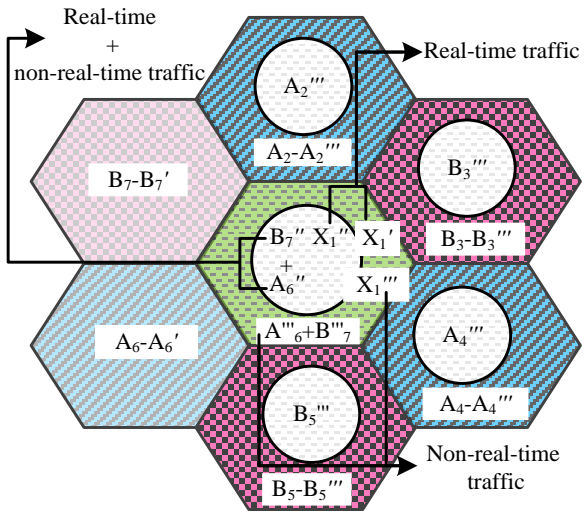

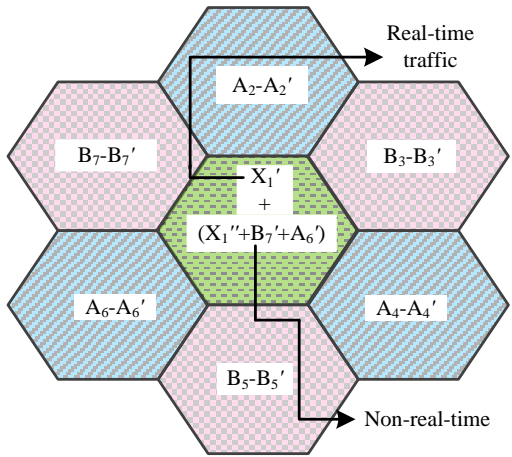

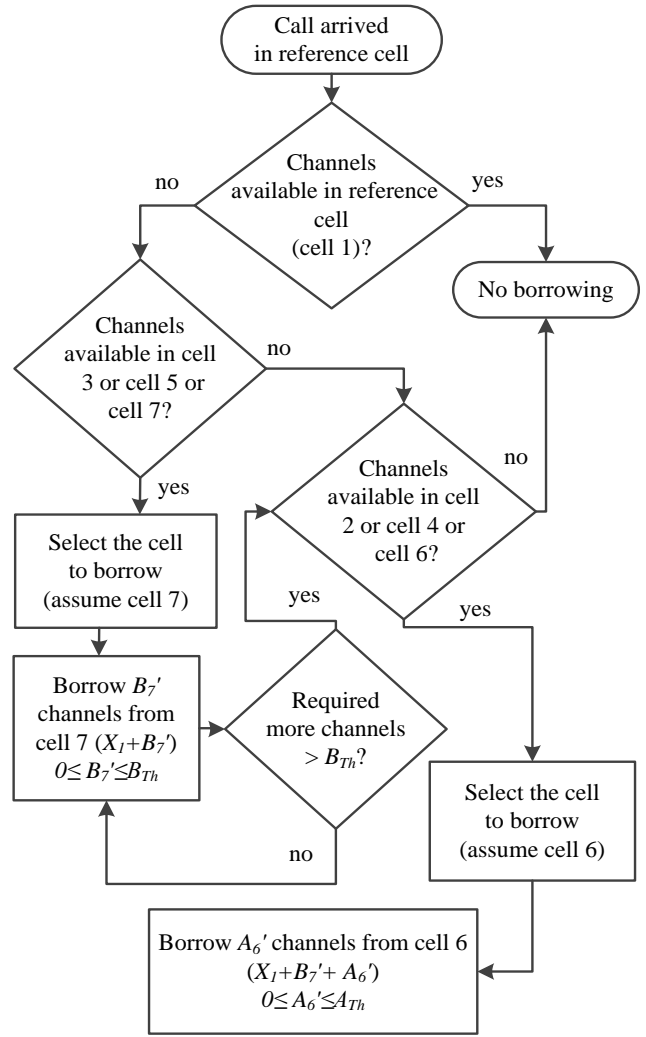

**Fig. 2.** Interference alleviation procedures giving priority to real-time traffic (a) bifurcation process of the reference cell and the adjacent cells (b) blockage of the interfering channels of the adjacent cells (c) bifurcation process of the reference cell and the adjacent cells when the blocked channels have to be activated.

**Fig. 3.** Dynamic channel assigning process.

The dynamic channel assigning process has been demonstrated in Fig. 3. When traffic arrives in reference cell, then it will search firstly among cell 3, cell 5 and cell 7. It can borrow up to the threshold ($A_{Th}$) from a cell of the group which has highest number of unoccupied channels. If more channels are required after borrowing from the group, then the cell will search among cell 2, cell 4 and cell 6 and borrow channels in a similar way up to threshold ($B_{Th}$). However, if there are no unoccupied channels in the group of cell 3, cell 5 and cell 7, then the reference cell will search in the other group.

The flow diagram of the proposed scheme shows the procedure in Fig. 4. When traffic arrives, the dynamic channel borrowing process is done. If the traffic is real-time, then it is assigned to the original channels of the reference cell which occupies the borrowed channels before assigning. Subsequently, the non-real-time traffic is assigned to the borrowed channels which were assigned to the original channels of the reference cell. If the traffic is non-real-time, then the interference is mitigated by the cell bifurcation process without channel assigning.

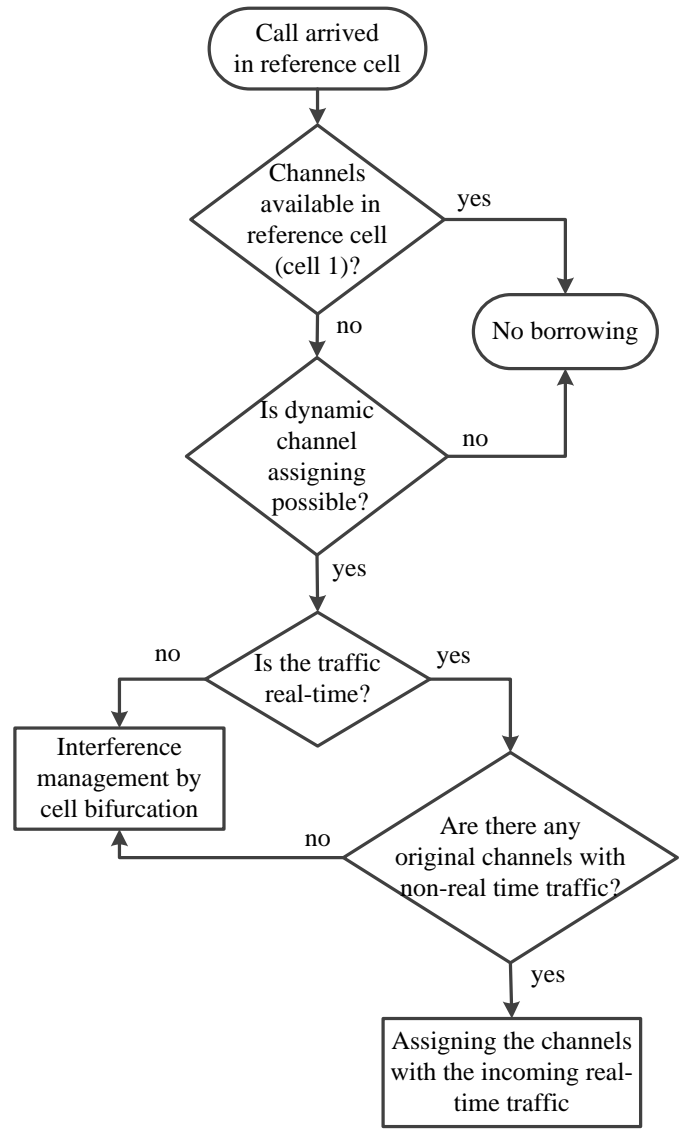

**Fig. 4.** Flow diagram of the proposed scheme for assigning the incoming real-time traffic to the original channels.

### III. OUTAGE PROBABILITY AND NETWORK CAPACITY ANALYSIS

Alleviation of interference and noise is one of the major challenges as this is responsible for degrading performances in the wireless networks. The consequences may end in the termination of an important call. In case of interference calculation, Okumura-Hata path loss model has been used [10]. From the model, we can write

$$L = 69.55 + 26.16 \log f_c - 13.82 \log h_b - a(h_m) \\ + (44.9 - 6.55 \log h_b) \log d + L_{ow} \quad (1)$$

$$a(h_m) = 1.1(\log f_c - 0.7)h_m - (1.56 \log f_c - 0.8) \quad (2)$$

where $d$ is termed as the distance between the BS and the user, $f_c$ is the centre frequency of the antenna, $h_m$ stands for the height of user (mobile antenna), $h_b$ is defined as the height of base station (BS) of the cell, $L_{ow}$ represents the penetration loss.

The received SINR level for the users can be expressed

$$SINR = \frac{S_0}{\sum_{m=1}^{P} I_m + \sum_{n=1}^{Q} I_n} \quad (3)$$

where $P$ and $Q$ are the number of interfering cells in 1st tier and 2nd tier, respectively, whereas $S_0$ is the signal power of the BS and $I$ represents the interference while the suffix represents the number of cell.

The capacity of wireless communication depends on the SINR level as Shannon capacity formula can be stated as

$$C = \log_2(1 + SINR) \quad [bps/Hz] \quad (4)$$

The outage probability can be expressed as

$$P_{out} = P_r(SINR < \gamma) \quad (5)$$

where γ is the threshold value of SINR level. We consider the value of γ in our performance analysis as 9 dB.

$$P_{out} = P_r\left(\frac{S_0}{\sum_{m=1}^{P} I_m + \sum_{n=1}^{Q} I_n} < \gamma\right) \quad (6)$$

Considering only the interfering channels of the cells, the outage probability can be expressed as

$$P_{out} = 1 - \prod_{i=1}^{P+Q} \exp\left(-\frac{\gamma}{S_o} I_i\right) \quad (6)$$

## IV. SIMULATION RESULTS

This section overviews the performances of the proposed scheme. We show the comparison of the performances in terms of SINR level, network capacity, and the outage probability. The performances of SINR level, outage probability, and the system capacity of the proposed scheme are compared to the approach where real-time traffic and non-real-time traffic are given equal priority without interference management. The outcomes of this section are demonstrated for real-time traffic and non-real-time traffic considering the proposed class based interference scheme and for the schemes which have not considered interference management. The result bears the novelty of the proposed scheme by ensuring the radio resource management with standard level of QoS. As the real-time traffic has to reassure more QoS than the non-real-time traffic, the scheme affords the preferred features for the QoS improvement for real-time traffic. However, only 1st and 2nd tier of the reference cell have been taken into consideration and threshold value of SINR (γ) is taken as 9 dB for the analyses of the performances. Table I typifies the fundamental parameters to analyses performances of the proposed scheme.

**Table I:** Parameter values used in the performance analysis.

| Parameter | Value |
|---|---|
| Original channel at each cell | 120 |
| Center frequency | 1800 MHz |
| Transmitted signal power by the BS | 1.50 kw |
| Cell radius | 1 km |
| Penetration loss | 10 dB |
| Threshold value of SINR (γ) | 9 dB |
| Height of the BS | 100 m |
| Height of the mobile antenna | 5 m |

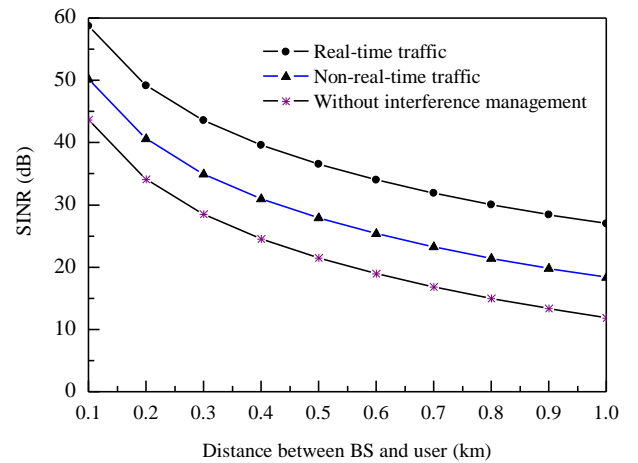

**Fig. 5.** Comparison of SINR levels

Figure 5 states the differences of SINR levels among the real-time traffic including those traffic which are given the original channels of the reference cell, the non-real-time traffic including those one which are assigned to the borrowed channels and the traffic without interference management. The result implies that the proposed interference management scheme escalates the SINR level for both real-time and non-real-time traffic which is very significant for uninterrupted communication. As the real-time traffic needs more QoS than the non- real-time traffic, the scheme provides the desired features for the QoS improvement for real-time traffic. Though the SINR level of the non-real-time traffic is less than the real-time traffic, it is more than that of the traffic without interference management. Moreover, the SINR level stays above the threshold level.

The comparison of the capacity between the proposed scheme and scheme without interference management is illustrated in Fig. 6. The capacity of the real-time traffic increases significantly which also ensures better QoS as well as the capacity of the non-real-time traffic stays in the acceptable range. Consequently, the proposed scheme shows a reliable data transmission rate than the scheme without interference management.

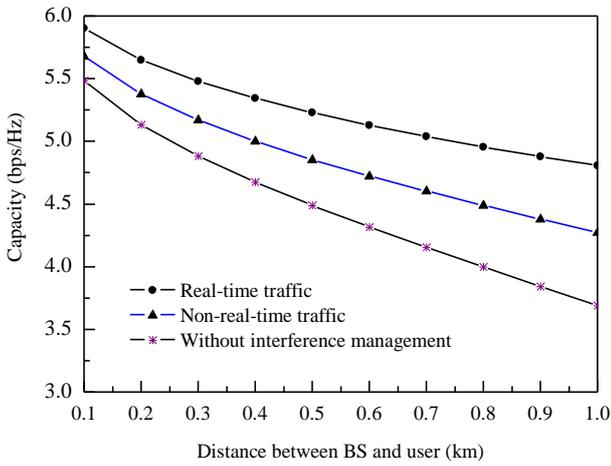

**Fig. 6.** Comparison of network capacity.

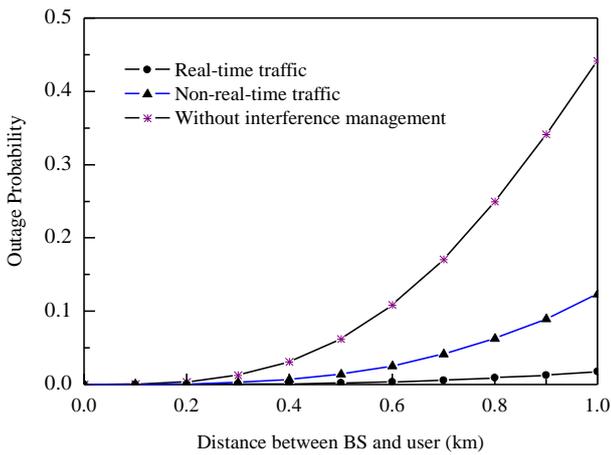

**Fig. 7.** Comparison of outage probability.

Figure 7 clarifies that the supremacy of the proposed scheme as the outage probability of the real-time traffic is profoundly smaller than the traffic without interference management. However, the outage probability of the non-real traffic is less the scheme without interference management. As the QoS of the non-real traffic is not very stringent, so the proposed scheme is very feasible. If the distance between BS and MS increases, the outage probability of the proposed scheme increases but the result remains in an acceptable range.

## V. CONCLUDING NOTES

Using limited same radio resource at the same time with insignificant interference for non-real-time traffic and utilization of non-interfering channels for real-time traffic makes the proposed model unique. To make it easy to comprehend, different illustrations are shown so that it can draw attention for future wireless networks. Utilization of interfering channels for non-real-time after interference management and non-interfering channels of the reference cell for real-time traffic makes the scheme quite operational. The scheme also provides outstanding performance in terms of system capacity, outage probability, and SINR level. Our future research includes priority scheme, multiclass traffic, intercellular interference mitigation, inter-carrier interference management in OFDMA, interference declination in multicellular networks. We are also interested to implement the proposed model for femtocellular networks.

## REFERENCES

[1] Ibrahim Habib, Mahmoud Sherif, Mahmoud Naghshineh, and Parviz Kermani, "An Adaptive Quality of Service Channel Borrowing Algorithm for Cellular Networks," *International Journal of Communication System*, vol. 16, no. 8, pp. 759–777, October 2003.

[2] Mostafa Zaman Chowdhury, Yeong Min Jang, and Zygmunt J. Haas, "Call Admission Control Based on Adaptive Bandwidth Allocation for Wireless Networks," *IEEE/KICS Journal of Communications and Networks*, vol. 15, no. 1, pp. 15-24, February 2013.

[3] Mohammad Arif Hossain, Shakil Ahmed, and Mostafa Zaman Chowdhury "Radio Resource Management for Dynamic Channel Borrowing Scheme in Wireless Networks," In Proceeding of *IEEE International Conference on Informatics, Electronics and Vision*, May 2014, pp. 1-5.

[4] Do Huu Tri, Vu Duy Loi, and Ha Manh Dao, "Improved Frequency Channel Borrowing and Locking Algorithm in Cellular Mobile Systems," In Proceeding of *IEEE International Conference on Advanced Communication Technology, February 2009, pp. 214 – 217.*

[5] Zuoying Xu, Pitu B. Mirchandani, and Susan H. Xu, "Virtually Fixed Channel Assignment in Cellular Mobile Networks with Recall and Handoffs," *Telecommunication Systems*, vol. 13, no. 2-4, pp. 413-439, July 2000.

[6] Geetali Vidyarthi, Alioune Ngom, and Ivan Stojmenovic´, "A Hybrid Channel Assignment Approach Using an Efficient Evolutionary Strategy in Wireless Mobile Networks," *IEEE Transactions on Vehicular Technology*, vol. 54, no. 5, pp. 1887-1895, September 2005.

[7] Shakil Ahmed, Mohammad Arif Hossain, and Mostafa Zaman Chowdhury, "Interference Declination for Dynamic Channel Borrowing Scheme in Wireless Networks," In Proceeding of *IEEE International Conference on Informatics, Electronics and Vision*, May 2014, pp. 1-5.

[8] Mostafa Zaman Chowdhury, Seung Que Lee, Byung Han Ru, Namhoon Park, and Yeong Min Jang, "Service Quality Improvement of Mobile Users in Vehicular Environment by Mobile Femtocell Network Deployment*,"* In Proceeding of *IEEE International Conference on ICT Convergence,* September 2011, pp. 194-198.

[9] Miguel López Benítez and Javier Gozalvez ,"Common Radio Resource Management Algorithms for Multimedia Heterogeneous Wireless Networks," *IEEE Transactions on Mobile Computing*, vol. 10, no 9, pp. 807-820, May 2012.

[10] Kaveg Pahlavan and Prasant Krishnamurthy, Principles of Wireless Networks, Prentice Hall PTR, New Jersey, 2002.